\documentclass[12pt]{article}
\usepackage{graphicx}
\usepackage{amssymb}
\begin{document}
\setcounter{section}{0}
\setcounter{equation}{0}
\noindent 
\begin{center}
{ \Large The influence of the hadronic interaction on the pionium wave functions}
\end{center}

\vspace{1.0cm}
\noindent 
\begin{center}
A. Gashi$^a$, G. Rasche$^a$$^\dagger$, W.S. Woolcock$^b$
\end{center}

\vspace{0.3cm}
\noindent
\begin{center}

  \noindent\textit{{$^{a}$Institut f\"{u}r Theoretische Physik der
  Universit\"{a}t,\\
        Winterthurerstrasse 190, CH-8057 Z\"{u}rich, Switzerland \\
        $^{b}$ Department of Theoretical Physics, IAS, \\
        The Australian National University, Canberra, ACT 0200, Australia}}
  \\
\end{center}

\noindent 
\vspace{0.7cm}

The influence of the hadronic $\pi\pi$ interaction on the wave functions of pionium
 (the Coulomb bound $\pi^+\pi^-$ system) at distances $\leq$ 10 fm is calculated for 
$s$-states with principal quantum numbers $n\leq 4$. Hadronic $\pi\pi$ potentials are used 
that reproduce the scattering phase shifts of two-loop chiral perturbation theory. 
The pionium wave functions $\psi_n (r)$ are obtained by integrating the coupled Schr\"odinger 
equations for the ($\pi^+\pi^-,\pi^0\pi^0$) system. We find that, for $r\leq 10$ fm, $n^{3/2} \psi_n (r)$ 
is practically independent of $n$. From this we conclude that the production rates of the $s$-states of 
pionium are proportional to $n^{-3}$, a result needed for the interpretation of the DIRAC experiment currently running at CERN.

\vspace{0.7cm}

\vspace{0.7cm}

\vspace{0.7cm}
\noindent $^\dagger$ Corresponding author. Electronic mail: rasche@physik.unizh.ch;
  Tel: +41 1 635 5810;  Fax: +41 1 635 5704

\newpage

The DIRAC experiment currently in progress at CERN[1], with the aim of measuring the lifetime $\tau_1$ of 
the 1$s$-state of pionium with an accuracy of 10 \%  , requires for the extraction of the lifetime an accurate knowledge 
of the production probability $w_n$ of the $ns$-states as a function of $n$. A proton beam of 24 GeV/c is incident on a target 
 (for example, titanium, nickel and platinum). Among the reaction products are many free $\pi^+\pi^-$ 
pairs and a much smaller number of $\pi^+\pi^-$ pairs bound into pionium, whose production occurs almost entirely 
in $ns$-states. These pionium atoms may decay into $\pi^0\pi^0$, with a lifetime $\tau_n$, or they may break 
up because of their interactions with the target atoms. The effect of the decay into $\gamma\gamma$ can be neglected here.  

The $\pi^+\pi^-$ pairs arising from the breakup of atoms nearly all have a relative momentum $\lesssim$ 3 MeV/c in their c.m. system. 
Observed pairs with a larger relative momentum are therefore almost all free pairs. This kinematical characteristic can be 
exploited to obtain experimentally for a given target the number of free pairs and the number of breakup pairs. By observing 
the number of breakup pairs from targets with different characteristics  it is possible to deduce the 
lifetime $\tau_1$.

This method of measuring  $\tau_1$ requires modelling of the processes of atom formation, decay (into  $\pi^0\pi^0$) and 
breakup. For this modelling it is necessary to know how $\tau_n$ and $w_n$ vary with $n$. To a high degree of accuracy, 
$\tau_n=n^3\tau_1$, independent of the details of the hadronic interaction. This result is easily obtained using the formalism of Ref.[2], which describes the decay process without  the use of wave functions. The behaviour of $w_n$ as a function of $n$ is the subject of the present paper.

For a collision process with pions among the final state particles, let $F({\bf P} ,{\bf q} )$ be the probability amplitude 
for the production of a free $\pi^+\pi^-$ pair with total momentum ${\bf P}$ and relative momentum ${\bf q}$ of the pair in 
its c.m. frame. The formation of $F$ involves an integration over the other variables that characterise the final state 
particles and a sum over all final states containing at least one $\pi^+$ and one $\pi^-$. Let $\tilde{F}({\bf P },{\bf x})$ be 
the Fourier transform of  $F({\bf P} ,{\bf q} )$ with respect to ${\bf q}$. Since the production of the pair is a hadronic 
process, $\tilde{F}({\bf P} ,{\bf x})$ is appreciably different from zero for $\mid\!\!{\bf x}\!\!\mid = r  \lesssim$ 10 fm at most.

According to Nemenov[3], where further details can be found, the probability amplitude for the production of a pionium atom 
in an $ns$-state with momentum  ${\bf P}$ is 
\[
\int d^3x\, \psi_n ({\bf x})^*\,\tilde{F}(\bf{P} ,\bf{x})\, ,
\]
where $\psi_n (\bf{x})$ is the pionium wave function in the rest frame of the atom. In fact  $\psi_n (\bf{x})$ is real and depends only on  $\mid\!\!{\bf x}\!\!\mid = r$. The probability for the production of a 
pionium atom with momentum $\bf{P}$ and principal quantum number $n$ is then
\begin{equation}
w_n = \mid \!\int d^3x\, \psi_n (r)\,\tilde{F}({\bf P} ,{\bf x})\! \mid^2.
\end{equation}
In the following we will show that, to a high degree of accuracy,
\begin{equation}
\psi_n(r) = n^{-3/2}\psi_1(r)\qquad \hbox {for} \qquad r \leq 10\, {\rm fm}\, .
\end{equation}
In view of our earlier observation on the behaviour of $\tilde{F}({\bf P },{\bf x})$ as a function of $r$, 
it follows from Eqs.(1) and (2) that
\begin{equation}
w_n = n^{-3}w_1
\end{equation}
to an even higher accuracy.

For a pure (i.e. point charge) Coulomb $\pi^+\pi^-$ interaction we know that, for small $r$,
\[
\psi_n(r) = n^{-3/2}\psi_1(r) (1 + O(r^2/B^2)).
\]
Since the Bohr radius $B$ for pionium is 387.5 fm, it follows that Eq.(2) holds in this case with an error of at most a few parts in $10^{4}$.
Adding the hadronic potential and taking account of the finite  extension of the charges results in a modification of $\psi_n (r)$ for r$ \lesssim$ 10 fm. Amirkhanov et al.[4] have made calculations of the modified  $\psi_n (r)$, but they ignore the 
possibility of the decay of the atom into  $\pi^0\pi^0$ and treat the problem as a single-channel one. They do 
not modify the Coulomb potential at small distances and they use a simple Yukawa-type $\pi^+\pi^-$ hadronic potential. 
They conclude that Eq.(2) continues to hold to a high level of accuracy.

Our work improves on Ref.[4] in three important respects. We have included the Coulomb potential for extended 
charges, exactly as in Ref.[5] (last paragraph of Section 4). We have used more realistic hadronic potentials and we 
have taken account of the presence of the  $\pi^0\pi^0$ channel and used coupled Schr\"odinger equations to model 
the ($\pi^+\pi^-,\pi^0\pi^0$) system. For the s-wave and for $2\mu_0<W<2\mu_c$ ($W$ is the total energy of the  $\pi\pi$ 
system in its c.m. frame and $\mu_c$ and $\mu_0$ are the masses of the charged and neutral pions) these may be written in the matrix form

\begin{equation}
(\mathbf{1}_2\frac{d^2}{dr^2}+\mathbf{Q}^2-\mbox{\boldmath$ \mu$}\mathbf{f}\mathbf{V}(r)) \mathbf{u}(r)=\mathbf{0}\, ,
\end{equation}
where 
\begin{equation}
\mbox{\boldmath$ \mu$} = \left( \begin{array}{cc}
\mu_c & 0  \\
0 & \mu_0 \end{array} \right) \, ,\,\,
\mathbf{Q}^2= \left( \begin{array}{cc}
-\kappa_c^2 & 0  \\
0 & q_0^2 \end{array} \right) \, , \,\, 
\mathbf{f}= \left( \begin{array}{cc}
f_c & 0  \\
0 & f_0 \end{array} \right) \, ,
\end{equation}
\begin{equation}
\mathbf{V}(r)={\mathbf{V}}^{em}(r)+\mathbf{V}^h(r) \, .
\end{equation}
Here 
\begin{equation}
\kappa_c^2=\mu_c^2-\frac{W^2}{4}\, , \,\,\,q_0^2=\frac{W^2}{4}-\mu_0^2 \,.
\end{equation}
Thus $q_0$ is the c.m. momentum in the $\pi^0\pi^0$ channel. The electromagnetic potential matrix is 
\[
\mathbf{V}^{em}=\left( \begin{array}{cc}
V^{em} & 0 \\
0 & 0 \end{array} \right)\, ,
\] 
where $V^{em}$ is the Coulomb potential between extended charges; there is no need to take account of the very small 
relativistic modification of the Coulomb potential or of vacuum polarisation. The hadronic potential matrix 
is assumed to be isospin invariant, in accordance with the arguments of Gasser and Leutwyler [6,7]. Thus
\begin{equation}
\mathbf{V}^h=\left( \begin{array}{cc}
\frac{2}{3}V^0+\frac{1}{3}V^2 & \frac{\sqrt{2}}{3}(V^2-V^0) \\
\frac{\sqrt{2}}{3}(V^2-V^0) & \frac{1}{3}V^0+\frac{2}{3}V^2 \end{array} \right)\, ,
\end{equation}
where $V^I$ is the hadronic potential for $l=0$ and total isospin $I$.
 
The diagonal matrix $\mathbf{f}$ has the nonzero components 
\begin{equation}
f_c=\frac{W^2-2\mu_c^2}{\mu_cW} \, , \, f_0=\frac{W^2-2\mu_0^2}{\mu_0W} \,.
\end{equation}
Its inclusion in the Schr\"{o}dinger equation using a potential has been extensively discussed in Refs.[8] and [9]. 
Since we will work very close to the $\pi^+\pi^-$ threshold, where the pionium states are located, it is sufficient to take 
\[
f_c=1 \, , \,\,\, f_0=\frac{4\mu_c^2-2\mu_0^2}{2\mu_0\mu_c} (>1) \,.
\]

The column vector $\mathbf{u}(r)$ is 
\begin{equation}
\mathbf{u}(r)= \left( \begin{array}{cc}
u_c(r)   \\
u_0(r) \end{array} \right) \, ,
\end{equation}
where
\begin{equation}
u_c(r)=\sqrt{4\pi}r\psi(r) \, , \, u_0(r)=rR_0(r) \, ,
\end{equation}
$\psi(r)$ being the $\pi^+\pi^-$ wave function and $R_0(r)$ the $\pi^0\pi^0$ radial wave function. For each value 
of $W < 2\mu_c$, there is a unique solution of Eq.(4) which satisfies 
\begin{equation}
\mathbf{u}(0)=\mathbf{0} \, ,
\end{equation}
as well as the asymptotic behaviour for $r \rightarrow \infty$
\begin{equation}
u_c(r)=A_c r^{\xi} e^{-\kappa_cr} \, , \,\,\, \xi=\frac{\alpha m_c}{\kappa_c}\, ,
\end{equation}
\begin{equation}
u_0(r)=A_0 \sin (q_0 r + \delta_{00})\, , \,\,\, A_0 > 0 \, ,
\end{equation} 
and in addition the normalisation condition 
\begin{equation}
\int_{0}^{\infty} dr (u_c(r))^2=1 \, .
\end{equation}

Here ${\delta}_{00}$ is the phase shift for elastic  $\pi^0\pi^0$ scattering and depends on $W$. If $W$ goes through energies 
$W_n$ that are close to the energies of the pure Coulomb bound states, $ \delta_{00}$ goes through an odd multiple of $90^0$. 
These energies are the bound state pionium energies with the influence of the hadronic $\pi\pi$ interaction taken into account; 
the functions  $\psi_n (r)$ at the energies  $W_n$ are the modified wave functions in which we are interested and which we have 
calculated for $n\leq 4$.

This completes the details of the calculation of the wave functions  $\psi_n (r)$, except for the hadronic potentials  
$V^{I}(r)$. We have constructed potentials that are independent of both the energy $W$ and the hadronic mass $\mu$, 
which reproduce very well the phase shifts given by two-loop chiral perturbation theory [10,11] for $\mu ={\mu} _0$ 
and $\mu ={\mu} _c$ via the Schr\"odinger equations

\begin{equation}
(\frac{d^2}{dr^2} + q^{2}_{c} - \mu_c\,f_c\,V^{I}(r))u^I(r,\mu_c) = 0 \, ,
\end{equation}

\begin{equation}
(\frac{d^2}{dr^2} + q^{2}_{0} - \mu_0\,f_0\,V^{I}(r))u^I(r,\mu_0) = 0 \, ,
\end{equation}
with $I = 0,2$. The  factors $f_c$ and $f_0$ are given in Eq.(9) and $q^{2}_{c}$ by
\[
q^{2}_{c} = \frac{W^2}{4} - {\mu}^{2}_{c}\qquad\hbox{,}\qquad W > 2\mu_c.
\]

The potentials $V^{I}(r)$ are of the same analytical form as those used for the calculation of the electromagnetic 
corrections in low energy pion-nucleon scattering [8,9]. In addition to a range parameter that was fixed at 1.5 fm, there are three 
further parameters that were varied to obtain the best fit. Full details of this parametric potential are given in Ref.[12]. 
The phase shifts were generated by integrating the regular solutions of Eqs.(16) and (17) outwards from the origin. 
The potentials  $V^{0}(r)$  and  $V^{2}(r)$ are plotted in Fig.1. We thus have captured part of the dynamical 
behaviour of chiral perturbation theory for $\pi\pi$ scattering by means of a simple model that uses potentials 
independent of the energy and the hadronic mass. This makes possible a reliable calculation of the pionium $ns$-state 
wave functions and therefore of the production rates $w_n$ as a function of $n$.

To show that $n^{3/2}\psi_{n}(r) \approx \psi_{1}(r)$ for $r \le 10$ fm for a range of values of $n$, we have calculated the quantity
\begin{equation}
d_n(r) = n^{3/2}\,\psi_{n}(r)/\psi_{1}(r) - 1,\,\,\,
r \le 10\,\, {\rm fm}\,,\,\,\,\,\,
n = 2,3,4.
\end{equation}
Fig.2 shows the result of this calculation. The values at $r=0$ are $d_2(0)=-0.0014$, $ d_3(0)=-0.0021$, $ d_2(0)=-0.0022$. As $r$  increases, $d_n(r)$ changes quickly at first and then becomes almost constant. The value of  $d_n(r)$ increases in magnitude as $n$ increases, but Fig.2 shows that Eq.(2) holds for $n = 2,3,4$ to better than 3 parts in $10^3$. The corresponding results for $n>4$ are not important, since only 2 \% of all pionium atoms are produced  in states with these higher principal quantum numbers. 

The results of Ref.[4] have the same sign as ours and show the same rapid change near $r=0$. At $r=8$ fm , Ref.[4] gives $d_2=-0.0012$, $d_3=-0.0010$ compared with our values of -0.0020 and -0.0026 respectively. Our more refined calculation yields results that are larger in magnitude but Eq.(3) still holds with an accuracy completely sufficient for use in modelling the DIRAC experiment [1].

{\bf Acknowledgements.} 
We wish to thank L. L. Nemenov for suggesting the calculation and for his comments on the text of the paper. We also thank G. Colangelo for providing us with the phase shifts predicted by two-loop chiral perturbation theory. We are indebted to the Swiss National Foundation for their financial support of the work.

\begin{figure}
  \begin{center}
    \includegraphics[height=0.75\textheight,angle=0]
      {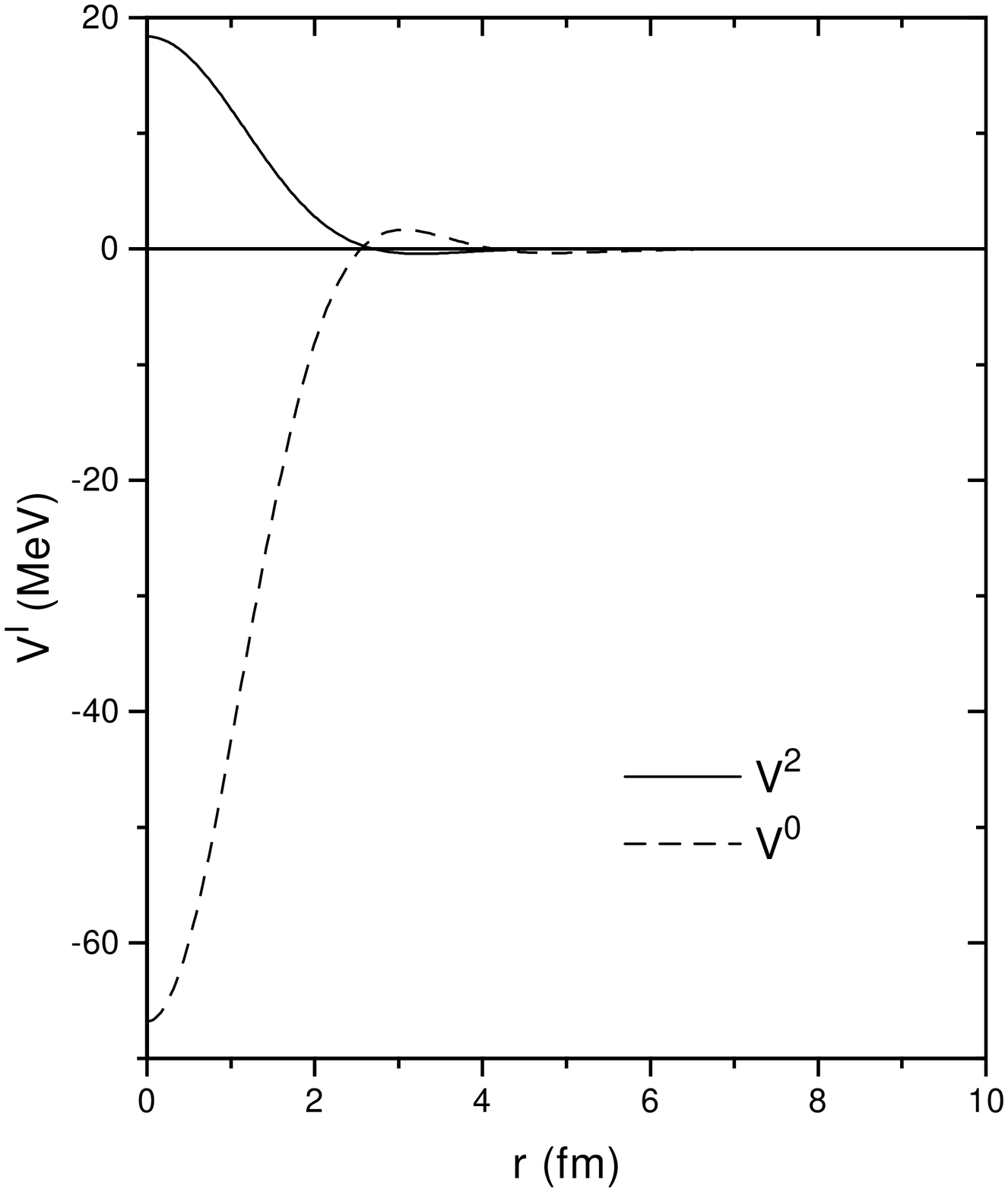}
      \caption{The potentials $V^0(r)$ and $V^2(r)$  that give the best fit to the two-loop 
  CHPT phase shifts of Ref.[10,11] for hadronic masses $\mu_c$ and $\mu_0$.}
    \end{center}
\end{figure}

\begin{figure}
  \begin{center}
    \includegraphics[height=0.75\textheight,angle=0]
      {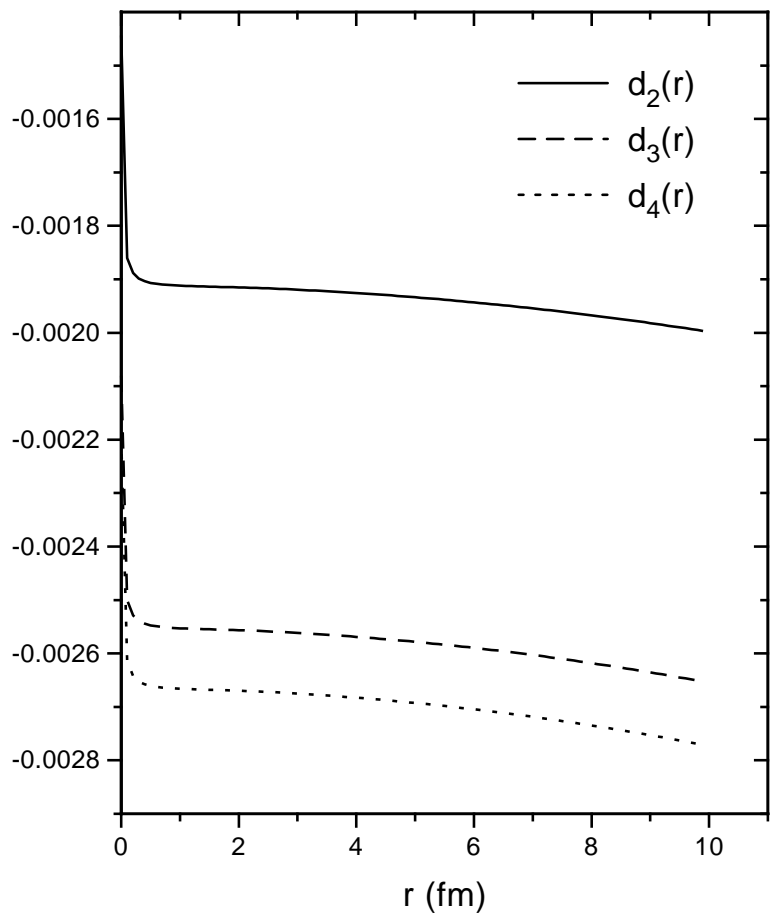}
      \caption{The functions $d_n(r)$ defined in Eq.(18) .}
    \end{center}
\end{figure}

\end{document}